# Laser tuning parameters and concentration retrieval technique for wavelength modulation spectroscopy based on the variable-radius search artificial bee colony algorithm

Tingting Zhang, Yongjie Sun, Pengpeng Wang, and Cunguang Zhu

*Abstract*—A novel wavelength modulation spectroscopy (WMS) laser tuning parameters and concentration retrieval technique based on the variable-radius-search artificial bee colony(VRS-ABC) algorithm is proposed. The technique imitates the foraging behavior of bees to achieve the retrieval of gas concentration and laser tuning parameters in a calibration-free WMS system. To address the problem that the basic artificial bee colony(ABC) algorithm tends to converge prematurely, we improve the search method of the scout bee. In contrast to prior concentration retrieval methods that utilized the Levenberg-Marquardt algorithm, the current technique exhibits a reduced dependence on the pre-characterization of laser parameters, leading to heightened precision and reliability in concentration retrieval. We validated the simulation with the VRS-ABC-based technique and the LM-based technique for the target gas $C_2H_2$. The simulation results show that the VRS-ABC-based technique performs better in terms of convergence speed and fitting accuracy, especially in the multi-parameter model without exact characterization.

*Index Terms*—wavelength modulated spectroscopy, calibration-free, spectral fitting, artificial bee colony.

## I. INTRODUCTION

Tunable diode laser absorption spectroscopy(TDLAS), a highly precise and sensitive technique for rapid detection of gas concentration, is extensively employed in various domains including, industrial gas detection, environmental air monitoring, and hazardous gas detection in agriculture and animal husbandry [1-8]. Wavelength modulation spectroscopy (WMS), which provides high sensitivity and stability, coupled with cost-effective instrumentation, stands out as one of the most widely adopted TDLAS techniques for in-situ measurements in harsh, high-temperature environments [9-16].

One major limitation of utilizing conventional WMS for temperature and concentration measurements in practical settings is the requirement to calibrate the WMS signals with a known mixture and condition. Such a calibration process can often be challenging and impractical for most field-deployable sensors and real-life environments. As a result, a number of researchers have proposed calibration-free methods for WMS, appropriate strategies include the utilization of the first harmonic ($1f$)-normalization and the application of multi-parameter fitting based on advanced Fourier series models towards the characterization of WMS lineshapes [17-19]. Using Fourier analysis, Stewart et al. [20] presented the theoretical basis for the application of RAM and phasor decomposition techniques, enabling precise line-shape recovery without the need for calibration, even under high modulation indexes or with high gas concentrations. Rieker et al. [21] utilized a normalization method by using the second harmonic ($2f$) normalized by the first harmonic ($1f$) to directly compare WMS measurements with WMS models to infer the temperature and concentration of gas, thereby eliminating the impact of light intensity fluctuations. Chao et al. [22] established an accurate model for scanning wavelength $1f$-normalized WMS-$2f$ of combustion gases NO sensors by characterizing intensity modulation and frequency modulation properties to achieve calibration-free measurements. A novel, practical calibration-free WMS scheme, utilizing WMS-$2f/1f$ spectral fitting, was recently introduced by Christopher et al. [23]. This method does not have limitations in the modulation amplitude, and does not require a theoretical description of WMS signals based on Taylor or Fourier series. Over the past few decades, the calibration-free methods for WMS have undergone significant development, where the Levenberg-Marquardt (LM) algorithm has gained prominent usage for spectral fitting aimed at inferring gas conditions. For example, Yang et al. [24] utilized the LM algorithm fitting the $2f$ spectrum. By using a reference signal to non-linearly fit the measured data with the LM algorithm, they were able to calculate the concentration of gas. This method enables the online measurement of low concentration methane. Cui et al. [25] have designed a CO sensor system with ppm-level

This work was supported by the Natural Science Foundation of China (Grant No. 61705080), the Promotive Research Fund for Excellent Young and Middle-Aged Scientists of Shandong Province (Grant No. ZR2016FB17), the Scientific Research Foundation of Liaocheng University (Grant No. 318012101), and the Startup Foundation for Advanced Talents of Liaocheng University (Grant No. 318052156 and 318052157). (Tingting Zhang and Yongjie Sun contributed equally to this work.) (*Corresponding author: Cunguang Zhu.*)

Tingting Zhang, Pengpeng Wang, and Cunguang Zhu are with the school of Physics Science and Information Technology, Liaocheng University, Liaocheng, 252000, China (e-mail: 2120110516@stu.lcu.edu.cn; wangpengpeng@lcu.edu.cn; cunguang_zhu@163.com).

Yongjie Sun is with the school of Physics and Technology, University of Jinan, Jinan, 250024, China (e-mail: 202021200721@stu.ujn.edu.cn).

sensitivity for analyzing the decomposition of sulfur hexafluoride in power systems. The system employs the LM algorithm to fit the 2f spectral information.

Although the LM algorithm effectively addresses most of the challenges [26-30] associated with gas property retrieval in calibration-free WMS systems, it is worth noting that its efficacy may be compromised when dealing with models that comprise multiple free parameters. With the increase in the number of free parameters, the dimensionality of the operation matrix in the LM algorithm undergoes a rise, eventually resulting in decreased accuracy and efficiency during model fitting. Therefore, caution should be exercised in the application of the LM algorithm under such circumstances. The challenges can be circumvented through the adoption of pre-characterization methodology of laser parameters in the construction of WMS models, which subsequently minimizes the number of free parameters. However, it may cause a series of secondary problems, such as measurement errors caused by the failure of the characterized values during the long operation of the instrument.

In this study, we propose a novel laser tuning parameters and concentration retrieval technique which utilizes the variable-radius-search artificial bee colony(VRS-ABC) algorithm for a calibration-free WMS system. Compared to the techniques based on the LM algorithm, the VRS-ABC-based technique exhibits relatively weak dependence on the pre-characterization of the laser tuning parameters. By enhancing the search capability of scout bees, this technique alleviates the issue of premature convergence commonly associated with the conventional ABC algorithm. Through simulation evaluation, we compare the performance of the VRS-ABC-based technique against the LM-based technique (Abbreviated as the VRS-ABC technique and LM technique) for concentration retrieval of $C_2H_2$ gas using a multi-parameter model without exact characterization. Our results demonstrate that the concentration retrieval technique based on the ABC algorithm excels in terms of convergence speed and fitting accuracy.

## II. THEORY AND METHODOLOGY

### A. Theory of WMS-2f/1f

In a WMS-2f/1f system, the laser is commonly driven by a low-frequency sinusoidal scanning current that is superimposed over a high-frequency sinusoidal modulation current. This results in a phase delay between the frequency and intensity of the emitted laser, which can be defined as

$$I_0(t) = \overline{I_0}[1 + i_1\cos(\omega t + \psi_1) + i_2\cos(2\omega t + \psi_2)] \quad (1)$$

$$v(t) = v_c + \Delta v \cdot \cos(\omega t) \quad (2)$$

where $\overline{I_0}(v_c)$ represents the mean laser intensity at the central laser frequency $v_c$, $i_1$ and $i_2$ denote the normalized linear and nonlinear intensity modulation (IM) depth, respectively, the modulation angular frequency $\omega$ is defined as $\omega=2\pi f_m$, where $f_m$ stands for laser modulation frequency (FM), $\varphi_1$ and $\varphi_2$ indicate the phase shift between the linear and nonlinear IM and FM, $\Delta v$ signifies the modulation depth of the laser frequency.

In accordance with the Beer-Lambert law, the laser beam attenuation through a gas cell filled with the absorbing gas, the transmitted laser intensity at frequency $v$ can be expressed as

$$I_t(t) = I_0(t) \cdot \exp[-\alpha(v(t))] \quad (3)$$

where $I_0(t)$ represents incident laser intensity, $\alpha(v(t))$ denotes the spectral absorbance, in the case of weak absorption ($\alpha < 0.05$).

$$I_t(t) = I_0(t) \cdot \exp[-\alpha(v(t))] \approx I_0(t) \cdot [1-\alpha(v(t))]$$
$$= I_0(t) \cdot [1-PS(T)CLg(v,v_0)] \quad (4)$$

where $S(T)$ represents the line strength of the absorption spectrum at temperature $T$, $P$ stands for total pressure, $C$ denotes the concentration of the gas being measured, $L$ represents the length of the optical range, $g(v, v_0)$ is the line shape function at the optical frequency $v$ of the absorption feature, $v_0$ denotes the line-center frequency of the absorption spectrum.

Under atmospheric pressure, the line shape function $g(v, v_0)$ can be represented by a Lorentzian line shape function

$$g(v,v_0) = \frac{2}{\pi \Delta v_c} \frac{1}{1+[x+m\cos(\omega t)]^2} \quad (5)$$

$$x = 2\frac{v_c - v_0}{\Delta v_c} \quad (6)$$

$$m = \frac{2\Delta v}{\Delta v_c} \quad (7)$$

where $\Delta v_c$ represents the full width at half-maximum, $x$ indicates the normalized frequency, $m$ is the modulation index.

Expansion of $\exp[-\alpha(v(t))]$ in a Fourier cosine series under sinusoidal injection current modulation of the laser is as follows:

$$\exp[-\alpha(v(t))] = \sum_{k=0}^{\infty} H_k(v_c, \Delta v) \cdot \cos(k\omega t) \quad (8)$$

here $H_k$ indicates the $k$th Fourier coefficient, which can be expressed as

$$H_0(v_c, \Delta v) = \frac{1}{2\pi}\int_{-\pi}^{\pi}\exp[-\alpha(v(t))]d\theta \quad (9)$$

$$H_k(v_c, \Delta v) = \frac{1}{\pi}\int_{-\pi}^{\pi}\exp[-\alpha(v(t))]\cdot\cos(k\theta)d\theta \quad (10)$$

The $I_t(t)$ signal is fed into the lock-in amplifier (LIA) and multiplied with a reference cosine wave and sine wave at $nf$ to extract the $X_{nf}$ and $Y_{nf}$ signals, respectively.

Subsequently, the X and Y components of the 1f signal are obtained after low-pass filtering as follows:

$$X_{1f} = \frac{\overline{I_0}}{2}\left[H_1 + i_0(H_0 + \frac{H_2}{2})\cos\varphi_1 + \frac{i_2}{2}(H_1 + H_3)\cos\varphi_2\right] \quad (11)$$

$$Y_{1f} = \frac{\overline{I_0}}{2}\left[i_0(H_0 - \frac{H_2}{2})\sin\varphi_1 + \frac{i_2}{2}(H_1 - H_3)\sin\varphi_2\right] \quad (12)$$

the X and Y components of the 2f signal can be obtained after low-pass filtering

$$X_{2f} = \frac{\overline{I_0}}{2}\left[H_2 + \frac{i_0}{2}(H_1+H_3)\cos\varphi_1 + i_2 H_0 \cos\varphi_2\right] \quad (13)$$

$$Y_{2f} = \frac{\overline{I_0}}{2}\left[\frac{i_0}{2}(H_1-H_3)\sin\varphi_1 + i_2 H_0 \sin\varphi_2\right] \quad (14)$$

after which the first harmonic normalized second harmonic $S_{2f/1f}$ are defined as

$$S_{2f/1f} = \sqrt{(\frac{X_{2f}}{X_{1f}})^2 + (\frac{Y_{2f}}{Y_{1f}})^2}$$

$$= \sqrt{(\frac{H_2 + \frac{i_1}{2}(H_1+H_3)\cos\psi_1 + i_2 H_0 \cos\psi_2}{H_1 + i_1(H_0 + \frac{H_2}{2})\cos\psi_1 + \frac{i_2}{2}(H_1+H_3)\cos\psi_2})^2 + (\frac{\frac{i_1}{2}(H_1-H_3)\sin\psi_1 + i_2 H_0 \sin\psi_2}{i_1(H_0-\frac{H_2}{2})\sin\psi_1 + \frac{i_2}{2}(H_1-H_3)\sin\psi_2})^2}$$

$$(15)$$

It is evident that the spectral model $S_{2f/1f}$ depends on a number of variables, including the modulation index $m$, gas concentration $C$, the depths of the linear and nonlinear IM $i_1$ and $i_2$, as well as the phases shift $\varphi_1$ and $\varphi_2$ between the linear and nonlinear IM and FM. This multi-parameter model poses a significant challenge for the LM technique, which we will effectively overcome using the VRS-ABC technique.

*B. VRS-ABC-based WMS concentration retrieval technique*

In this paper, we developed a VRS-ABC-based concentration retrieval technique that simulates the foraging behavior of a bee colony to perform calibration-free measurements of gas concentration and laser tuning parameters.

The optimization process of the VRS-ABC algorithm consists of the behavior of three types of bees: employed bees, onlooker bees and scout bees. Among them, employed bees are in charge of gathering and preserving data on reliable nectar sources. Onlooker bees keep updating their sources of nectar based on the information provided by the employed bees. Scout bees will evaluate if the current nectar sources are ideal before looking for new ones. The three types of bees complement each other and dynamically switch roles to find the optimal solution to the problem eventually.

The VRS-ABC technique flowchart is shown in Fig. 1. The specific processes are as follows. The free parameters of the fitting procedure may include the modulation index $m$, gas concentration $C$, the depths of the linear and nonlinear IM $i_1$ and $i_2$, as well as the phases shift $\varphi_1$ and $\varphi_2$ between the linear and nonlinear IM and FM. We denote the free parameters by the vector $\beta$ that defines the current position of the nectar source.

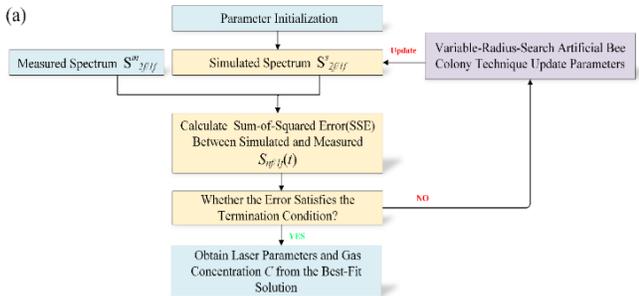
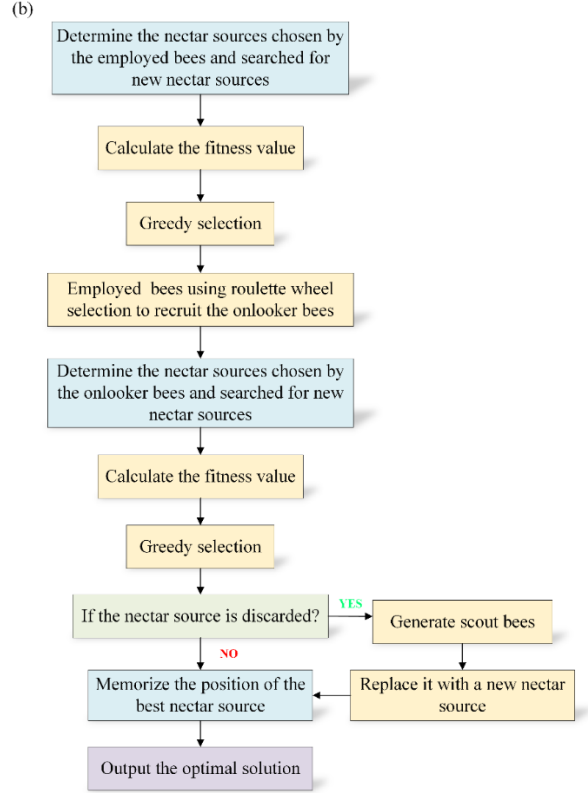

**Fig. 1.** (a) The flowchart of the VRS-ABC technique. (b) The VRS-ABC technique update parameters.

Step 1: Parameter initialization - The vectors $\beta_i$ generated randomly within the range of the boundaries of the free parameters, $i= 1, 2, …, W$.

Step 2: Acquisition of simulated spectra - The simulated spectra $f_i(x_k)$ are obtained by substituting the initialized parameters $\beta_i$ into Eq. (15), $x_k$ is the normalized frequency, $k= 1, 2, 3, …, N$, $N$ is the total number of sampling points in the simulated spectra.

Step 3: Parameters updating - The fitness is used to evaluate the amount of nectar from each nectar source, which is proportionally correlated with each other. The fitness value $fitness_i$ is defined as follows:

$$fitness_i = 1/(1+F_i) \quad (16)$$

$$\begin{aligned} &Given\ F_i : R^D \to R \\ &Find \\ &z_k = y_k - f_i(x_k, \beta_i) \\ &F(\beta_i) = \sum_{k=1}^{N} z_k^2 \\ &End \end{aligned} \quad (17)$$

where $F_i$ is the objective function, $R^D$ represents the $D$-dimensional solution space, $D$ is dependent on the number of free parameters, $y_k$ is measured spectra, $z_k$ is the residual between the $k$th sampling points of measured spectra and simulated spectral, and the objective optimization function $F(\beta_i)$ is the sum of squares of all residuals.

Step 4: Judgment - If the convergence between measured spectra $S^m_{2f/1f}$ and simulated spectral $S^s_{2f/1f}$ does not satisfy the termination condition of the optimization, then continue to Step 5; Otherwise, perform Step 6.

Step 5: Bee movement - Each employed bee can only be associated with a single nectar source, meaning the number of nectar sources equals the number of employed bees. The process of seeking a new nectar source by employed bee $x_i$ corresponding to the $i$th nectar source is defined as:

$$v_{id} = x_{id} + \lambda_{id}(x_{id} - x_{pd}) \tag{18}$$

where $v_{id}$ is the possible solution for the nectar source, $\lambda$ is a uniformly distributed real random number within the range of [-1,1], $d$ indicates the $d$th free parameters, $d= 1, 2, \ldots, D, p \in \{1, 2\ldots, W\}$ is a randomly chosen index that has to be different from $i$.

For a current position $x_i$ and its neighbor $v_i$, a greedy selection is applied. The algorithm chooses the better one as the next position based on their fitness values. If the fitness of $v_i$ is higher than that of $x_i$, the algorithm updates the current position to $v_i$. If the fitness of $x_i$ cannot improve, the algorithm increments a counter by 1; otherwise, the counter is reset to 0.

Onlooker bees searched for new nectar sources using a roulette wheel selection scheme in which the size of each piece was proportional to the fitness value:

$$p_i = \frac{fitness_i}{\sum_{i=1}^{W} fitness_i} \tag{19}$$

where $fitness_i$ is the fitness value of $v_i$. Clearly, the greater the fitness value of a given nectar source, the higher its likelihood of being selected as a potential solution. Once the onlooker bee has selected a nectar source, it will proceed to search for it, according to Eq. (17).

After completing their searches, the employed bees and onlooker bees evaluate the fitness value of a nectar source and discard it if it doesn't improve within the given limit parameter. Consequently, the bee responsible for that source becomes a scout bee. In the basic ABC algorithm, the scout bee is responsible for randomly searching for new nectar sources in the following manner:

$$x_{id} = x_d^{min} + \eta(x_d^{max} - x_d^{min}) \tag{20}$$

where $\eta \in [0,1]$ is a random value, $x_d^{min}$ and $x_d^{max}$ are the lower and upper bounds of the $D$-dimensional parameters.

The well-balanced between local and global exploration activities assist in alleviating both stagnation and premature convergence of the algorithm. However, in the basic ABC algorithm, the weak global search capability and being entrapped in a local optimum too early is still the main problem. Such a situation could result in significant inaccuracies during the retrieval of concentration and laser parameters. Introducing a variable radius search approach [31] to the scout bee can effectively resolve the inherent deficiencies of the basic ABC algorithm.

In each iteration, they modify the search radius for potential candidates using a larger radius, followed by a gradual reduction in radius as the convergence of the process approaches. The search formula for the scout bee is updated to

$$v_{id} = x_{id} + \eta[\omega_{max} - \frac{iteration}{MaxIt}(\omega_{max} - \omega_{min})]x_{id} \tag{21}$$

where $v_{ij}$ represents a new viable solution generated by a scout bee, which is altered from the current position of a discarded nectar source $x_{ij}$, $MaxIt$ and $iteration$ is the maximum and the current number of iterations, respectively. The parameters $\omega_{max}$ and $\omega_{min}$ specify the upper and lower bounds, respectively, on the percentage change allowed in the position of the scout bee. These parameters are set to fixed values of 1 and 0.2, respectively. Using these selected parameter values, the position adjustment of the scout bee declines linearly from 100 percent to 20 percent concerning its current position in each round of experimentation.

The scout bee's solution will deviate significantly from the optimal solution during the initial iteration. The global search capability of the algorithm is strong at this point. With the increase in the number of iterations, the scout bee's step size will gradually decrease, and the local search of the algorithm begins to dominate.

Step 6: Termination condition - If the convergence between measured spectra $S_{2f/1f}{}^m$ and simulated spectra $S_{2f/1f}{}^s$ satisfies the termination condition of the optimization, the value of best-fit parameters is output at this point, and laser parameters and gas concentration $C$ are obtained.

TABLE I
SUMMARY OF SPECTRAL PARAMETERS

| Symbol | Quantity | Value |
| --- | --- | --- |
|  | isotopologue | $^{12}C_2H_2$ |
| $v_c$ | line-center frequency of transition | 6523.8792 cm$^{-1}$ |
| $\Delta v_c/2$ | half width at half-maximum | 0.0777 cm$^{-1}$ |
| $S$ | line strength | 1.035× 10$^{-20}$ cm/mol |
| $T$ | arbitrary temperature | 296 K |
| $P$ | total pressure | 1 atm |
| $L$ | absorption path length | 20 cm |

III. SIMULATION

The feasibility of the VRS-ABC-based WMS concentration retrieval technique is verified in the following. We decided to carry out the validation by simulation in MATLAB R2019b platform to avoid the characterization errors introduced by the metrology instruments affecting the evaluation of the algorithm performance. In the same circumstances, we compared the effects of the basic ABC, VRS-ABC and LM techniques on concentration retrieval.

We have selected the spectral line P(13) of acetylene gas at a wavelength of 1532.83 nm as our target line. Given the values of the spectral parameters in Table 1. The measured spectra that should have been collected in the experiment were replaced by the virtual measurement data set calculated by Eq. (19):

$$y_n = [(x_1, y_1),(x_2, y_2),...,(x_{4000}, y_{4000})] \tag{22}$$

Fig. 2 shows the first harmonic normalized second harmonic $S_{2f/1f}$ acetylene images at different concentrations.

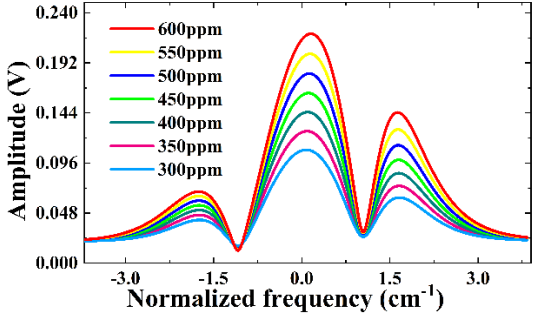

**Fig. 2.** Images of acetylene $S_{2f/1f}$ under different concentrations.

To verify the effectiveness of the improvement to the algorithm, we conducted a comparative analysis of performance between the VRS-ABC algorithm and the basic ABC algorithm. During the thirty seconds fitting process, various parameters such as the modulation index ($m$), gas concentration ($C$), linear and nonlinear IM depth ($i_1$ and $i_2$), and the phase shift ($\varphi_1$ and $\varphi_2$) are considered free parameters. Fig. 3 shows the predictions of the VRS-ABC technique for different concentrations of acetylene, showing an excellent linear dependence between the expected concentration value and the predicted concentration. We analyzed the predicted concentration at 450 ppm, as depicted in Fig. 4. Furthermore, a comprehensive systematic error analysis was performed by comparing the statistical indicator of relative error (RE), as presented in Fig. 5. It is apparent that during the initial iteration, the VRS-ABC technique experiences a considerable error. However, as the number of iterations increases, the algorithm's error diminishes significantly during the later stages. The REs of the final VRS-ABC technique are smaller than those of the basic ABC technique. This reflects the dynamic adjustment of the VRS-ABC technique to the search range with the number of iterations, which greatly improves the algorithm's ability to search the space. The spectrum simulated using the predictions of the VRS-ABC technique showed good agreement with the virtual spectrum, as shown in Fig. 6. The concentration and laser tuning parameters predicted by the VRS-ABC technique are shown in Table 2. The predicted values are statistically consistent with the expected values. Notably, the REs of all laser parameters is below 5%, with the RE for concentration $C$ being a mere 0.11%. This demonstrates the high reliability and accuracy of the VRS-ABC technique, providing further evidence of the advantages of this approach.

After verifying the performance advantages of the VRS-ABC technique in a multi-parameter model with laser tuning parameters and gas concentration, a comparison with the LM technique was carried out. Specifically, we employ the LM technique to predict the simulated spectrum while maintaining a convergence time of thirty seconds. As shown in Fig. 7, some details are not satisfactory. Fig. 8 displays the concentration curves predicted by both VRS-ABC and LM methodologies over time, where the LM approach exhibits significantly lower efficiency. The concentration and laser parameters, as predicted by the LM technique, are presented in Table 3. It is clear that the predicted values for both concentration $C$ and nonlinear IM depth $i_2$ exhibit significant deviation from the expected values. And they all have REs of more than 10%. The impact of this on the gas concentration retrieval from the multi-parameter WMS model is self-evident.

According to the above simulation results, the VRS-ABC-based concentration retrieval technique has a faster fitting rate and higher fitting accuracy than the LM technique. This can also be demonstrated for accurately predicting laser tuning parameters and concentrations. In conclusion, the satisfactory multi-parameter optimization of the VRS-ABC technique can eliminate the dependence on pre-characterization and prevent measurement errors caused by pre-characterization failure.

TABLE II
THE CONCENTRATION AND LASER PARAMETERS PREDICTED BY THE VRS-ABC TECHNIQUE WITH A CONVERGENCE TIME OF 30 S

| Free parameters | Expected value | Predicted by the VRS-ABC technique | Relative Errors |
| --- | --- | --- | --- |
| $m$[cm$^{-1}$] | 1.500 | 1.484 | 1.07% |
| $c$[ppmv] | 450.0 | 450.5 | 0.11% |
| $i_1$[pm/mA] | 0.150 | 0.1435 | 4.33% |
| $i_2$[pm/mA] | 0.003 | 0.0029 | 3.33% |
| $\varphi_1$[π] | 0.600 | 0.605 | 0.83% |
| $\varphi_2$[π] | 0.500 | 0.497 | 0.60% |

TABLE III
THE CONCENTRATION AND LASER PARAMETERS PREDICTED BY THE LM TECHNIQUE WITH A CONVERGENCE TIME OF 30 S

| Free parameters | Expected value | Predicted by the VRS-ABC technique | Relative Errors |
| --- | --- | --- | --- |
| $m$[cm$^{-1}$] | 1.500 | 1.415 | 5.67% |
| $c$[ppmv] | 450.0 | 535.5 | 19.0% |
| $i_1$[pm/mA] | 0.150 | 0.155 | 3.33% |
| $i_2$[pm/mA] | 0.003 | 0.0034 | 13.3% |
| $\varphi_1$[π] | 0.600 | 0.6210 | 3.50% |
| $\varphi_2$[π] | 0.500 | 0.533 | 6.60% |

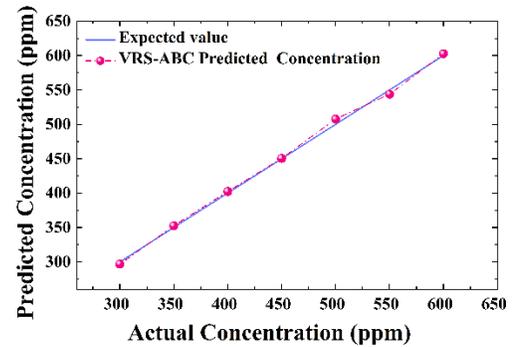

**Fig. 3.** Relationship between expected and predicted concentrations.

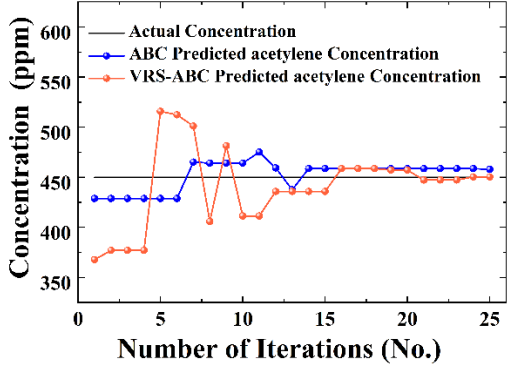

**Fig. 4.** Analysis of the predicted concentrations.

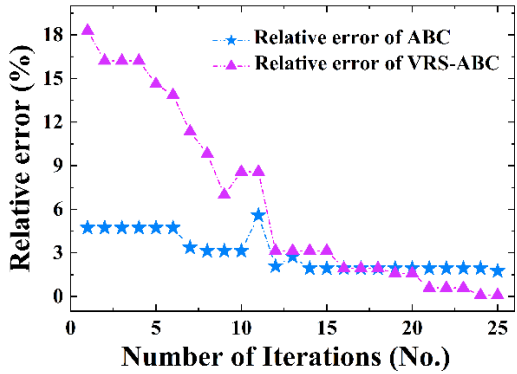

**Fig. 5.** Error analysis of predicted concentration by the ABC and VRS-ABC techniques.

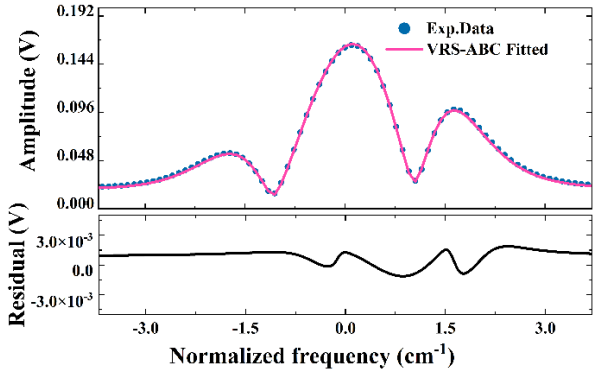

**Fig. 6.** Fitting effect of the VRS-ABC technique with a convergence time of 30 s.

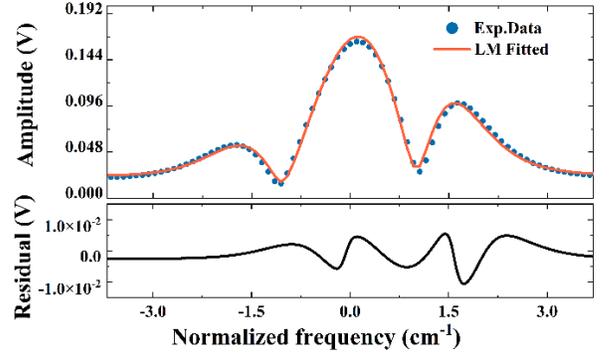

**Fig. 7.** Fitting effect of the LM technique with a convergence time of 30 s.

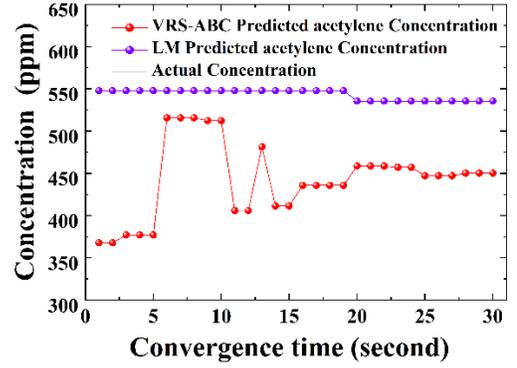

**Fig. 8.** Analysis of the predicted concentration changes over time by LM technique

## IV. CONCLUSION

In this paper, a novel WMS laser tuning parameters and concentration retrieval technique based on the VRS-ABC algorithm is proposed. The technique is used in a calibration-free WMS system to achieve the retrieval of gas concentration and laser tuning parameters. By improving the search method of the scout bee, we solve the problem that the basic ABC algorithm tends to converge prematurely. In contrast to prior concentration retrieval methods that utilized the LM algorithm, the VRS-ABC technique exhibits a reduced dependence on the pre-characterization of laser parameters. We validated the simulation with the VRS-ABC technique and the LM technique for the target gas $C_2H_2$. The results show that the REs of the concentration $C$ and nonlinear IM depth $i_2$ predicted by the LM technique exceed 10%. However, the REs of all laser parameters predicted by the VRS-ABC technique is below 5%, with the RE for concentration $C$ being a mere 0.11%. The technique can get rid of the gas concentration measurements from the dependence on pre-characterization of laser parameters and also prevent measurement errors caused by pre-characterization failure.

REFERENCES

[1] H. Li, G. B. Rieker, X. Liu, J. B. Jeffries and R. K. Hanson, "Extension of wavelength-modulation spectroscopy to large modulation depth for

[1] diode laser absorption measurements in high-pressure gases," *Appl. Opt*, vol. 45, no. 5, pp. 1052–1061, 2006.
[2] A. Hangauer, J. Chen, R. Strzoda, M. Ortsiefer and M. C. Amann, "Wavelength modulation spectroscopy with a widely tunable InP-based 2.3 μm vertical-cavity surface-emitting laser," *Opt. Lett*, vol. 33, no. 14, pp. 1566–1568, 2008.
[3] K. Sun, X. Chao, R. Sur, J. B. Jeffries and R. K. Hanson "Wavelength modulation diode laser absorption spectroscopy for high-pressure gas sensing," *Appl. Phys. B*, vol. 110, pp. 497–508, 2013.
[4] L. Dong, F. K. Tittel, C. Li, N. P. Sanchez, H. Wu, C Zheng and R. J. Griffin "Compact TDLAS based sensor design using interband cascade lasers for mid-IR trace gas sensing," *Opt. Express*, vol. 24, no. 6, pp. A528–A535, 2016.
[5] H. Lu, C. Zheng, L. Zhang, Z. Liu, F. Song, X. Li, Y. Zhang and Y. Wang, "A remote sensor system based on TDLAS technique for ammonia leakage monitoring," *Sensors*, vol. 21, no. 7, pp. 2448, 2021.
[6] Z. Peng, Y. Ding, L. Che, X. Li and K. Zheng "Calibration-free wavelength modulated TDLAS under high absorbance conditions," *Opt. Express*, vol. 19, no. 23, pp. 23104–23110, 2011.
[7] C. Liu, L. Xu, J. Chen, Z. Cao, Y. Lin and W. Cai "Development of a fan-beam TDLAS-based tomographic sensor for rapid imaging of temperature and gas concentration," *Opt. Express*, vol. 23, no. 17, pp. 22494–22511, 2015.
[8] O. Witzel, A. Klein, C. Meffert, S. Wagner, S. Kaiser, C. Schulz and V. Ebert, "VCSEL-based, high-speed, in situ TDLAS for in-cylinder water vapor measurements in IC engines" *Opt. Express*, vol. 21, no. 17, pp. 19951–19965, 2013.
[9] K. Sun, X. Chao, R. Sur, C. S. Goldenstein, J. B. Jeffries and R. K. Hanson, "Analysis of calibration-free wavelength-scanned wavelength modulation spectroscopy for practical gas sensing using tunable diode lasers," *Meas. Sci. Technol*, vol. 24, no. 12, pp. 15203, 2013.
[10] C. Li, L. Dong, C. Zheng and F. K. Tittel "Compact TDLAS based optical sensor for ppb-level ethane detection by use of a 3.34 μm room-temperature CW interband cascade laser," *Sensor. Actuat. B. Chem*, vol. 232, pp. 188–194, 2016.
[11] H. Li, A. Farooq, J. B. Jeffries and R. K. Hanson, "Near-infrared diode laser absorption sensor for rapid measurements of temperature and water vapor in a shock tube," *Appl. Phys. B*, vol. 89, pp. 407–416, 2007.
[12] M. Raza, K. Xu, Z. Lu and W. Ren "Simultaneous methane and acetylene detection using frequency-division multiplexed laser absorption spectroscopy," *Opt. Laser. Technol*, vol. 154, pp. 108285, 2022.
[13] G. B. Rieker, H. Li, X. Liu, J. B. Jeffries, R. K. Hanson, M. G. Allen, S. D. Wehe, P. A. Mulhall and H. S. Kindle, "A diode laser sensor for rapid, sensitive measurements of gas temperature and water vapour concentration at high temperatures and pressures," *Meas. Sci. Technol*, vol. 18, no. 5, pp. 1195, 2007.
[14] G. B. Rieker, H. Li, X. Liu, J. T. C. Liu, J. B. Jeffries, R. K. Hanson, M. G. Allen, S. D. Wehe, P. A. Mulhall, H. S. Kindl, A. Kakuho, K. R. Sholes, T. Matsuura, S. Takatani, "Rapid measurements of temperature and H2O concentration in IC engines with a spark plug-mounted diode laser sensor," *P. Combust. Inst*, vol. 31, no. 2, pp. 3041–3049, 2007.
[15] C. Li, C. Zheng, L. Dong, W. Ye, F. K. Tittel and Y. Wang, "Ppb-level mid-infrared ethane detection based on three measurement schemes using a 3.34-μm continuous-wave interband cascade laser," *Appl. Phys. B*, vol. 122, pp. 1–13, 2016.
[16] J. Norooz Oliaee, N. A. Sabourin, S. A. Festa-Bianchet, J. A. Gupta, M. R. Johnson, K. A. Thomson, G. J. Smallwood and P. Lobo, "Development of a sub-ppb resolution methane sensor using a GaSb-based DFB diode laser near 3270 nm for fugitive emission measurement," *Acs. Sensors*, vol. 7, no. 2, pp. 564–572, 2022.
[17] A. Upadhyay and A. L. Chakraborty, "Calibration-free 2f WMS with in situ real-time laser characterization and 2f RAM nulling," *Opt. Lett*, vol. 40, no. 17, pp. 4086–4089, 2015.
[18] N. Liu, L. Xu, S. Zhou, L. Zhang and J. Li, "Soil respiration analysis using a mid-infrared quantum cascade laser and calibration-free WMS-based dual-gas sensor," *Analyst*, vol. 146, no. 12, pp. 3841–3851, 2021.
[19] A. J. McGettrick, K. Duffin, W. Johnstone, G. Stewart and D. G. Moodie, "Tunable diode laser spectroscopy with wavelength modulation: A phasor decomposition method for calibration-free measurements of gas concentration and pressure," *J. Lightwave. Technol*, vol. 26, no. 4, pp. 432–440, 2008.
[20] G. Stewart, W. Johnstone, J. R. P. Bain, K. Ruxton and K. Duffin "Recovery of absolute gas absorption line shapes using tunable diode laser spectroscopy with wavelength modulation—Part I: Theoretical analysis," *J. Lightwave. Technol*, vol. 29, no. 6, pp. 811–821, 2011.
[21] G. B. Rieker, J. B. Jeffries and R. K. Hanson, "Calibration-free wavelength-modulation spectroscopy for measurements of gas temperature and concentration in harsh environments," *Appl. Opt*, vol. 48, no. 29, pp. 5546–5560, 2009.
[22] X. Chao, J. B. Jeffries and R. K Hanson, "Wavelength-modulation-spectroscopy for real-time, in situ NO detection in combustion gases with a 5.2 μm quantum-cascade laser," *Appl. Phys. B*, vol. 106, pp. 987–997, 2012.
[23] C. S. Goldenstein, C. L. Strand, I. A. Schultz, K. Sun, J. B. Jeffries and R. K. Hanson, "Fitting of calibration-free scanned-wavelength-modulation spectroscopy spectra for determination of gas properties and absorption lineshapes," *Appl. Opt*, vol. 53, no. 3, pp. 356–367, 2014.
[24] R. Yang and Y. Zhang, "A method of low concentration methane measurement in tunable diode laser absorption spectroscopy and Levenberg-Marquardt algorithm," *Optik*, vol. 224, pp. 165657–165671, 2020.
[25] R. Cui, L. Dong, H. Wu, S. Li, L. Zhang, W. Ma, W. Yin, L. Xiao, S. Jia and F. K. Tittel, "Highly sensitive and selective CO sensor using a 2.33 μm diode laser and wavelength modulation spectroscopy," *Opt. Express*, vol. 26, no.19, pp. 24318–24328, 2018.
[26] W. Wei, J. Chang, Q. Huang, Q. Wang, Y. Liu and Z. Qin, "Water vapor concentration measurements using TDALS with wavelength modulation spectroscopy at varying pressures," *Sensor. Rev*, vol. 37, no. 2, pp. 172–179, 2017.
[27] S. Wagner, M. Klein, T. Kathrotia, U. Riedel, T. Kissel A. Dreizler and V. Ebert, "In situ TDLAS measurement of absolute acetylene concentration profiles in a non-premixed laminar counter-flow flame," *Appl. Phys. B*, vol. 107, pp. 585–589, 2012.
[28] W. Y. Peng, C. L. Strand and R. K. Hanson, "Analysis of laser absorption gas sensors employing scanned-wavelength modulation spectroscopy with 1f-phase detection," *Appl. Phys. B*, vol. 126, no. 1, pp. 17, 2020.
[29] J. Li, Z. Peng and Y. Ding, "Wavelength modulation-direct absorption spectroscopy combined with improved experimental strategy for measuring spectroscopic parameters of H2O transitions near 1.39 μm," *Opt. Lasers. Eng*, vol. 126, pp. 105875, 2020.
[30] Y. Sun, P. Wang, T. Zhang, K. Li, F. Peng and C. Zhu, "Principle and Performance Analysis of the Levenberg–Marquardt Algorithm in WMS Spectral Line Fitting," *Photonics*, vol. 9, no. 12, pp. 999, 2022.
[31] A. Banharnsakun, T. Achalakul and B Sirinaovakul, "The best-so-far selection in artificial bee colony algorithm," *Appl. Soft. Comput*, vol. 11, no. 2, pp. 2888–2901, 2011.

**Tingting Zhang** was born in Dezhou, Shandong, China, in 1997. She is currently pursuing the M.S. degree at the school of physics science and information technology with Liaocheng University. Her current research interests include optical gas sensing devices, optical fiber sensor fabrication and intelligent algorithms.

**Yongjie Sun** was born in Zibao, Shandong, China, in 1997. He is currently pursuing the M.S. degree in physics and technology with University of Jinan. His current research interests include optical fiber sensor fabrication and intelligent algorithms.

**Pengpeng Wang** was born in Jinan, Shandong, China, in 1985. She received the Ph.D. degree from Shandong University in 2014. She is currently with the School of Physics Science and Information Technology, Liaocheng University, Liaocheng, China. Her current research interests include optical fiber sensors and fiber lasers.

**Cunguang Zhu** was born in Liaocheng, Shandong, China, in 1985. He received the Ph.D. degree in optoelectronic engineering from Shandong University, Jinan, China, in 2015. He is currently with the School of Physics Science and Information Technology, Liaocheng University, Liaocheng. His current research interests include optical gas sensing devices, optical fiber sensor fabrication, and engineering applications.